\documentclass[twocolumn,superscriptaddress,showpacs,prl,floatfix]{revtex4}
\usepackage{graphicx,epsfig,psfrag}
\usepackage{amsmath}

\newcommand{\be}{\begin{equation}}
\newcommand{\ee}{\end{equation}}
\newcommand{\bea}{\begin{eqnarray}}
\newcommand{\eea}{\end{eqnarray}}

\begin{document}
\title{
Real and complex random neutrino mass matrices and $\theta_{13}$ 
\boldmath 
  
\unboldmath}
 \author{  Janusz Gluza}
\affiliation{Institute of Physics, University of Silesia, 
            Uniwersytecka 4, PL-40007 Katowice, Poland }
 \author{  Robert Szafron}
\affiliation{Institute of Physics, University of Silesia, 
            Uniwersytecka 4, PL-40007 Katowice, Poland }
 
\begin{abstract} 
Recently  it has been shown that one of the basic parameters of the neutrino sector, so called $\theta_{13}$ angle is very small, but quite probably non-zero.  
We argue that the small value of $\theta_{13}$ can still be reproduced easily by a wide spectrum of randomly generated models of neutrino masses. For that we consider real and complex neutrino mass matrices, also including sterile neutrinos. A qualitative difference between results for real and complex mass matrices in the region of small $\theta_{13}$ values is observed.
We show that statistically the present experimental data prefers  random models of neutrino masses with sterile neutrinos.
 \end{abstract}

\maketitle 
\allowdisplaybreaks

Since their first appearance in modern science neutrinos 
have played an important part in our understanding of the laws of particle
physics.
To see how
important the weak interaction of neutrinos is in Nature 
let’s just mention the mechanism in which the Sun is shining \cite{bah}. 
It is also very well known that they influence  evolution of the whole Universe as their tiny masses have an impact on the dynamics of the expansion of the Universe 
\cite{cosmol1,cosmol2}.
Without any
doubt the investigation of the properties of these particles can reveal many interesting, hidden until now physical phenomena or explain many hypothetical
ideas.
 
The theory of neutrino oscillations  requires nonzero neutrino masses as well as nonzero 
neutrino mixing angles.  

The T2K Collaboration recently announced that a new  
measurement~\cite{Abe:2011sj} has yielded a nonzero $\theta_{13}$ for CP conserving case with $\delta_{CP}=0$ (results at 90\% 
confidence level), i.e.
\begin{eqnarray}
0.03 \leq &\sin^2 2 \theta_{13} &\leq 0.28, \; \mbox{\rm normal mass hierarchy,}\\
0.04 \leq & \sin^2 2 \theta_{13} &\leq 0.34,\; \mbox{\rm inverted mass hierarchy.}
\label{th13h}
\end{eqnarray} 

The Minos collaboration published similar results few days later \cite{Adamson:2011qu}. These results with nonzero $\theta_{13}$ are of primary importance for the future of particle physics \cite{moh,giunti}. 

Present global fits of the data give \cite{Fogli:2011qn} (at 1$\;\sigma$ C.L., \footnote{Fits in \cite{Schwetz:2011zk} differs slightly, but the exact values of parameters are not the main issue here.}) 
\begin{eqnarray}
 7.05 \cdot 10^{-5} ~{\rm eV}^2   \leq & \Delta m^2_{21}  &\leq 8.34 \cdot 10^{-5} ~{\rm eV}^2, \label{eq1}\\
 2.07 \cdot 10^{-3} ~{\rm eV}^2  \leq & \Delta m^2_{32} &\leq 2.75 \cdot 10^{-3} 
~{\rm eV}^2,  \\
 0.39 \leq & \sin^2 \theta_{23} & \leq 0.5,  \\ 
0.291 \leq & \sin^2 \theta_{12} & \leq 0.324, \label{eq1a}  \\ 
  0.008 \leq & \sin^2 \theta_{13}& \leq 0.036 \label{eq2}. 
\end{eqnarray}

There are many theoretical models which tune the space of possible neutrino parameters to get agreement with 
these experimental data. The new data from the T2K collaboration generated a new challenge for that.  
Frequently the pattern of a flavour symmetry is invoked, for  recent theoretical adjustments of the nonzero $\theta_{13}$ mixing angle, see e.g. \cite{Ma:2011yi}.
Here we face the problem differently asking how natural are small, nonzero values of $\theta_{13}$.
Naturalness means that the neutrino mass matrices are generated randomly. Apart from neutrino physics, see e.g. \cite{Hall:1999sn,Dziewit:2006cg}, random matrices 
came to be an eminent tool used in many different fields of science, including day-to-day life problems 
\cite{math}. 

Let us first consider a symmetric, three dimensional matrix with random elements in the
range $[-1,1]$. 
Physically this case realizes Majorana type of neutrinos. Similarly, we consider also  Dirac neutrinos (generated by general real random matrices)
and so-called see-saw neutrinos with masses $m_\nu$ defined through relation $m_D^T M_R^{-1} m_D$ (here again matrices $m_D$ and $m_R$ are generated randomly in the
range $[-1,1]$). Next, we apply the singular value decomposition theorem \cite{nash} to calculate the mixing matrix $U$  which diagonalizes the random matrices we started from. Using the standard parametrisation,  
 we are able to link experimental mixing angles to the elements of matrix $U$, e.g. $\sin^2{\theta_{13}} = |U_{13}|^2$.  

\begin{figure}[tb]
\epsfig{figure=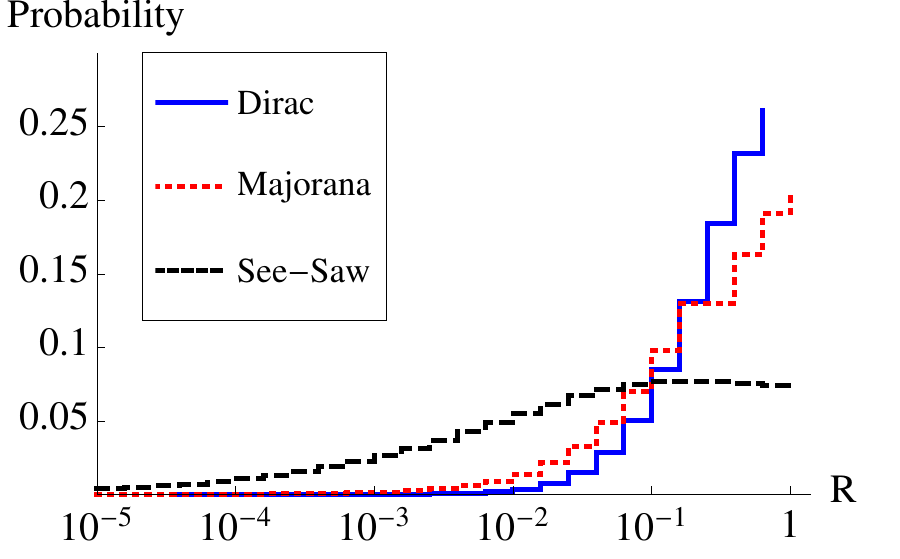,height=5cm}
\caption{\label{Rfig}  Frequency probability for $R$ parameter obtained from $10^7$ randomly generated  three dimensional different types of matrices. 
No constraints given  by Eqs.\ref{eq1}-\ref{eq2} are applied.}
\end{figure}

In Fig.\ref{Rfig} we recover one of the results discussed already in \cite{Hall:1999sn}, see Fig.1. Here $R=\Delta m_{21}^2/\Delta m_{32}^2$ is the ratio of the smallest to the next smallest values of differences  of squared neutrino masses, $m_1<m_2<m_3$. Both in Fig.\ref{Rfig} and next plots results are generated based on $10^7$ random mass matrices. Hall et al. approach has an advantage of pure simplicity. A modified versions have been considered in many papers, see e.g. \cite{Vissani:2001im, hep-ph/0210342, hep-ph/0301050,arXiv:1104.0602, arXiv:1108.0964}.

Let us note that the see-saw neutrino masses are trivially not well defined if elements of $M_R$ approach zero in relation $m_D^T M_R^{-1} m_D$, which can happen as elements of $M_D$ and $M_R$  are generated in the $[-1,1]$ range. Thus we will  consider only Dirac and Majorana type of neutrinos.
To disentangle these two types of neutrinos is of great importance for neutrino physics \cite{Kayser:1989iu}. If we connect the mechanism of generation of their tiny masses with some heavy states, this issue starts to be important for high energy colliders like $e^+e^-$ \cite{Gluza:1996bz,Aguila:2005pf}, $e^-e^-$ \cite{Gluza:1995ix} 
and LHC \cite{Aguila:2007em}. 

We can extend this discussion to the case where elements of the neutrino mass matrix  involve complex phases
(random numbers with modulus
in the interval $[0, 1]$ and a phase in a range  $[0; 2 \pi]$). It might seem that real and imaginary matrices should produce qualitatively the same results: situation differs only as an additional degree of freedom (phase) enters in each entry of the mass matrix. However, this is not true, as in fact they differ.
This change does not affect the shapes of the eigenvalues distributions in a significant way but other observables distributions like elements of the mixing matrix may change.
This is exactly the case for $\theta_{13}$.

\begin{figure}[tb]
\epsfig{figure=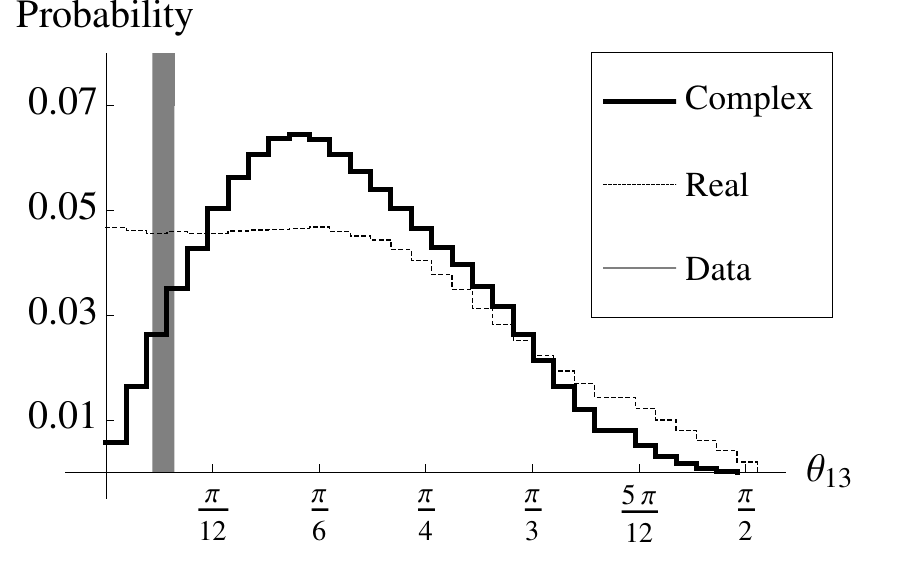,height=5.5cm}
\caption{\label{reim}  Frequency probability for $\theta_{13}$ obtained from $10^7$ randomly generated real and imaginary three dimensional symmetric matrices
(Majorana neutrinos). No constraints given  by Eqs.\ref{eq1}-\ref{eq1a} are applied.
Gray vertical band represents experimental values of $\theta_{13}$, Eq.\ref{eq2} at the 1$\sigma$ level.
}
\end{figure}
 
 \begin{figure}[tb]
\epsfig{figure=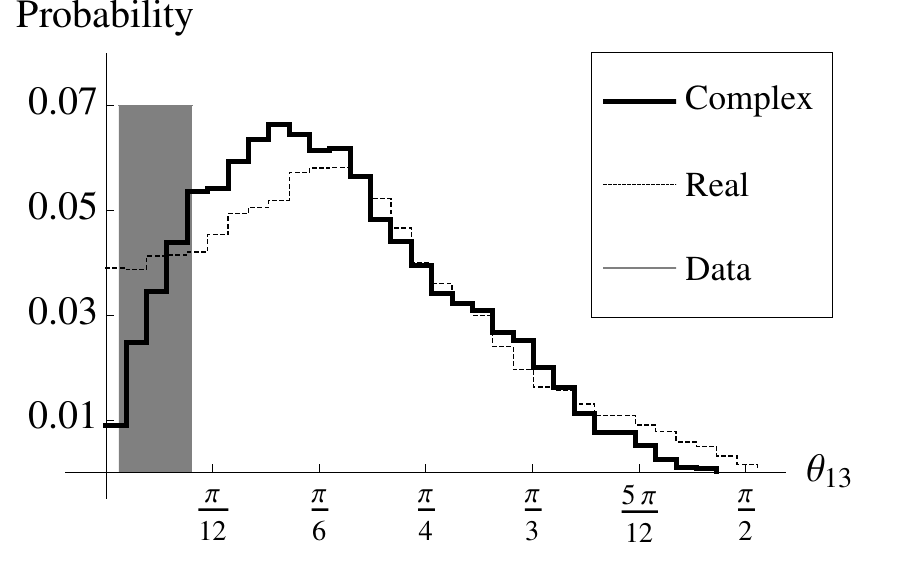,height=5.5cm}
\caption{  Frequency probability for $\theta_{13}$ obtained from $10^7$ randomly generated real and imaginary three dimensional symmetric matrices
(Majorana neutrinos). Constraints by Eqs.\ref{eq1}-\ref{eq1a}  are applied at the $3 \sigma$ level.
Gray vertical band represents experimental values of $\theta_{13}$, Eq.\ref{eq2} at the 3$\sigma$ level.}
\label{reimconstr}
\end{figure}

 Fig.\ref{reim} shows the results of numerical predictions for $\theta_{13}$, additional constraints given by  Eqs.\ref{eq1}-\ref{eq1a} are not applied.
We can see that the real and imaginary random matrices  behave  qualitatively differently in vicinity of small $\theta_{13}$
and real matrices fit better. 
That frequency probability of random complex matrices tends to zero with $\theta_{13} \to 0$ can be understood in the following way. For real matrices CP complex phase is zero or $\pi$ and CP is not violated when $\theta_{13}=0$ and therefore there is no direct restriction on values of $\theta_{13}$. Reversing 
 the argument, if  CP symmetry is violated (which is the case for randomly generated complex mass matrices) then $\theta_{13}$ must be different from zero.  
 We found that frequency probability for the physical CP violating phase $\delta$ which emerges from  randomly generated mass matrices is constant (so this is a trivial plot and we do not show it here. From pure mathematics (a probability measure) follows that in such case a probability to get $\delta$ equal to zero or $\pi$ is 
 zero, so probability distributions should tend to zero for $\theta_{13}\to 0$, what can be seen on histograms, see solid line in Fig.~\ref{reim}.
If Eqs.\ref{eq1}-\ref{eq1a} are applied, a number of mass matrices which cover the experimentally interesting values of   $\theta_{13}$  decreases.
For instance, using $1 \sigma$ cuts in Eqs.\ref{eq1}-\ref{eq1a},    only  4886 real and 4686 complex matrices remain out of initial $10^7$ random matrices.
  Using $3 \sigma$ cuts 
we are left with 31853 real and 29141 complex matrices, see Fig.\ref{reimconstr}. 
Comparing between Dirac and Majorana cases  more mass matrices in
Majorana case survive the cuts.
The main conclusion remains however the same as in the case where no cuts Eqs.\ref{eq1}-\ref{eq1a} are applied: there are more matrices which reproduce experimental data for real matrices and complex matrices do not condense in 
the experimentally interesting region.

It is interesting that the last experimental data implies the presence of a fourth sterile neutrino \cite{Giunti:2011ht,Giunti:2011gz,Yasuda:2011np,Giunti:2011hn}.
 Usually the so called 3+1 and 3+2 models of sterile neutrinos are considered~\footnote{For the see-saw mechanism, there are additional right-handed singlets with a much higher mass scale. But it is not the only possible scenario. Still, light sterile neutrinos are possible. 
 That is why we extend the anarchy neutrino mass model of Haba et al. directly to higher dimensions. Moreover, this can be a kind  of 
an effective mass matrix which originates from more complicated mechanisms.}.
 We then inspect the mass matrices of dimensions four and five. 
 Here we focus on an element $|U_{13}|^2$  and  define an effective angle such that $\sin^2{\theta_{13}}= |U_{13}|^2$,  so we can compare the results obtained from
 models with different number of neutrinos.

\begin{figure}[tb]
\epsfig{figure=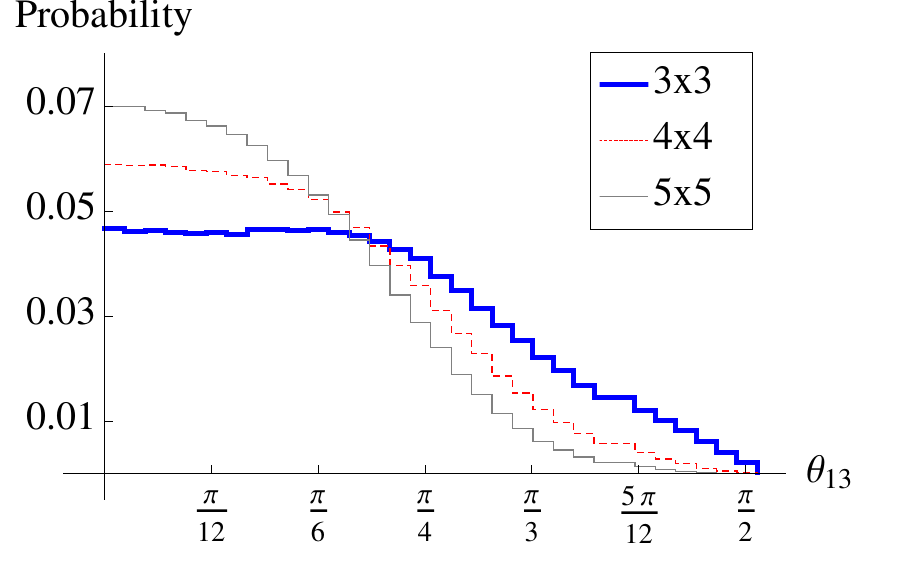,height=5cm}
\caption{Shifts in the direction of small values of $\theta_{13}$ with increasing number of sterile neutrinos.  No constraints given  by Eqs.\ref{eq1}-\ref{eq1a} are applied.}
\label{fig345}
\end{figure}
  
In a case of more than three dimensional mass matrices   relations among the elements of the mixing matrix and experimentally defined mixing angles are more involved, therefore we will consider this case without  cuts driven by Eqs.\ref{eq1}-\ref{eq2} and without a discussion of the impact of additional 
mixing matrix parameters of four and five dimensional mass matrices. We proceed as in the previous
case and generate random matrices and then calculate
the mixing matrix.
In Fig.\ref{fig345}   we show results for Majorana  (symmetric) real matrices of different dimensions. 
We observe that the more  neutrinos we have the bigger  number
of random matrices reproduce the experimental value of the   $\theta_{13}$ parameter.
Similar results give Dirac neutrinos.

We have no clear explanation why the probability of getting small values for $\theta_{13}$ grows with dimensionality of matrices. There is a mathematical law of  Wigner
\cite{wig} connected with real symmetric random matrices which states that their eigenvalues accumulate around zero (Wigner's semicircle law \cite{wolf} is
exact in the limit of infinite dimensions of matrices). To our knowledge, there is no relation connecting the  distributions of the values of elements of eigenvectors with the dimensionality of real symmetric matrices. 

Nonetheless, let us try to understand the result we obtained. 
In the anarchical model of neutrino masses, when no cuts on parameters are  
applied, mixing angles have very similar probability  
distributions. This is because every element of the mass matrix has  
exactly the same probability distribution.  Therefore, if the dimension  
of the mass matrix grows, the unitarity  of the mixing matrix  
constrains sums of squared modulus of mixing matrix  
elements, and  an average value of each of elements of the mixing matrix will  decrease  
with increasing dimensionality. This forces the probability distribution to  
take higher values for small angles as the dimension grows, 
a kind of pattern obtained in Fig.\ref{fig345}.

\begin{figure}[tb]
\epsfig{figure=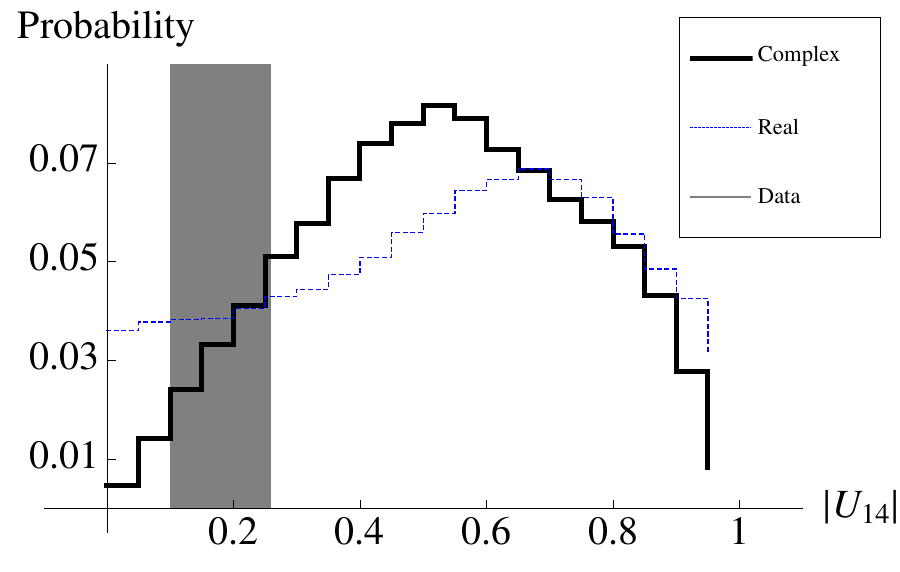,height=5cm}
\caption{Probability distributions for $|U_{14}|$ elements of the neutrino mixing matrix when the constraint Eq.\ref{eq2} for  $\theta_{13}$ 
is applied. The gray vertical band stands for $3\sigma$ values which follow from  global fits \cite{giunti3p1b}.}
\label{fig5}
\end{figure}

\begin{figure}[tb]
\epsfig{figure=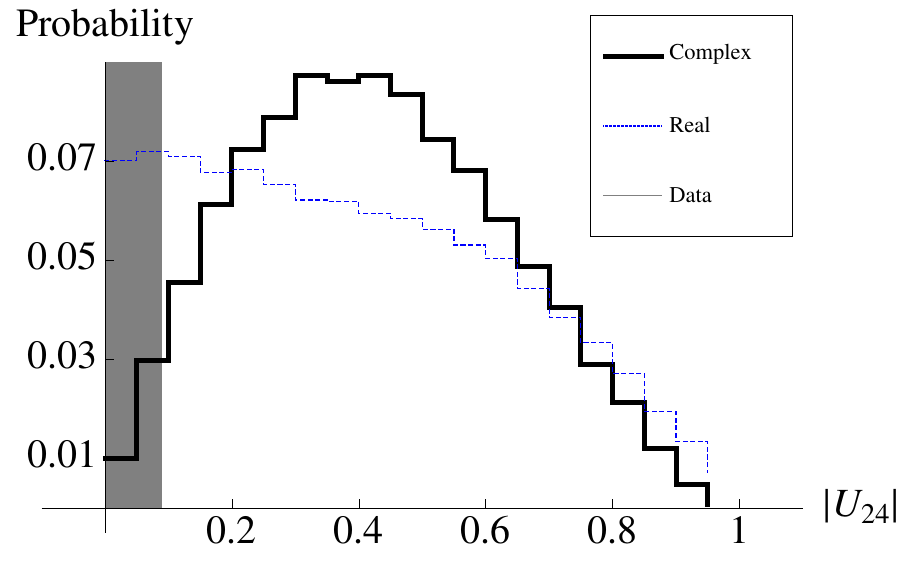,height=5cm}
\caption{Probability distributions for $|U_{24}|$ elements of the neutrino mixing matrix when the constraint Eq.\ref{eq2} for  $\theta_{13}$ 
is applied. The gray vertical band stands for upper bounds derived by MINOS \cite{min}.}
\label{fig6}
\end{figure}

We can make a step further and consider for the first time  implications of nonzero $\theta_{13}$ on values of elements $|U_{14}|$ and
$|U_{24}|$ in the $3+1$ sterile model when the constraint Eq.\ref{eq2} for  $\theta_{13}$ 
is applied. (Without cuts, probability distributions for   $|U_{14}|$ and  $|U_{24}|$ are practically the same as 
for $\theta_{13}$ in Fig.\ref{reim}, as the just discussed equal-distribution 
argument suggests.)
It is interesting that analyses of recent experiments predict non-zero values for them \cite{giunti3p1a,giunti3p1b}.
In Fig.~\ref{fig5} and Fig.~\ref{fig6} the plots are given for real and complex neutrino Majorana mass matrices (Dirac patterns are very similar). We can see that probability distributions for both real and complex cases reach a maximum at larger values of $|U_{14}|$
($|U_{14}|_{max} \simeq 0.7$ for a real case, $|U_{14}|_{max} \simeq 0.58$ for a complex case)
when compared with $|U_{24}|$ ($|U_{24}|_{max} \simeq 0.1$, real case;  $|U_{14}|_{max} \simeq 0.4$, complex case). This is in agreement with fits discussed in \cite{giunti3p1b} ($|U_{\mu 4}|^2 < |U_{e4}|^2$ for both low and high energy neutrino experimental fits).
In Fig.\ref{fig5} we can see that the most frequent values of $|U_{14}|$ do not coincide with its experimentally preferable region, though in this experimental region they can also be substantial.

More involved analyses including other experimental constraints and various mass hierarchies in $3+1$ and $3+2$ models go beyond 
this Brief Report and are left for a future work.

In conclusion,  though
random matrices can not solve fundamental problems in neutrino physics, they generate intriguing hints on the nature of neutrino mass matrices.

\vspace{1cm}
\begin{acknowledgments}
We would like to thank Marek Gluza and Radomir Sevillano for useful discussions and careful reading of the manuscript. Work supported in part by the Research Executive Agency (REA) of the European Union under the
Grant Agreement number PITN-GA-2010-264564 (LHCPhenoNet), 
by the Polish Ministry of Science under
grant No. N N202 064936 and National Science Centre.
\end{acknowledgments}
\providecommand{\href}[2]{#2}

\providecommand{\href}[2]{#2}

\end{document}